\documentclass[12pt]{article}

\usepackage{scicite}

\usepackage{times}

\newenvironment{sciabstract}{%
\begin{quote} \bf}
{\end{quote}}

\newcounter{lastnote}

\title{Orbital Motion in the Radio Galaxy 3C 66B:
        Evidence for a Supermassive Black Hole Binary}

\author
{Hiroshi Sudou,$^{1\ast}$ Satoru Iguchi,$^{2}$ Yasuhiro Murata,$^{3}$
Yoshiaki Taniguchi,$^{1}$\\
\\
\normalsize{$^{1}$Astronomical Institute, Graduate School of Science,}\\
\normalsize{Tohoku University, Sendai 980-8578, Japan}\\
\normalsize{$^{2}$National Astronomical Observatory of Japan,}\\
\normalsize{Mitaka, Tokyo 181-8588, Japan}\\
\normalsize{$^{3}$Institute of Space and Astronautical Science, }\\
\normalsize{Sagamihara, Kanagawa 229-8510, Japan}\\
\\
\normalsize{$^\ast$To whom correspondence should be addressed; E-mail:  sudou@astr.tohoku.ac.jp.}
}

\date{}

\begin{document} 


\baselineskip24pt


\maketitle


\begin{sciabstract}
Supermassive black hole binaries may exist in the centers of
active galactic nuclei like quasars and radio galaxies and mergers 
between galaxies may 
result in the formation of 
supermassive binaries during the course of galactic evolution. 
Using the very-long-baseline interferometer, we imaged  the radio galaxy 3C 66B at radio frequencies 
and  found that the unresolved radio core of 3C 66B
shows well-defined elliptical motions with a period of 1.05 $\pm$ 0.03  years,
which provides a direct detection of a supermassive black hole binary.
\end{sciabstract}


The presence of a supermassive black hole binary (or a supermassive
binary; SMB) in active galactic nuclei (AGNs)
has been suggested mainly by 
periodic optical and radio outbursts ({\it 1,2}),
wiggled patterns of compact radio jets (i.e., the indication of precession
motions of the radio jets) ({\it 3,4}), and
X-shaped morphology of radio lobes ({\it 5}). 
AGNs with those characteristics tend to be associated with strong radio
jets and thus 
their host galaxies are often giant elliptical galaxies ({\it 6}).
If they were made from mergers between or among nucleated galaxies ({\it 
7,8}),
they may contain two or more black holes in their central regions
because progenitor galaxies may contain a supermassive black hole in
their nuclei ({\it 9,10}). Captured black holes will be settled
under dynamical friction in the core of the merged stellar system and then 
will form a SMB ({\it 11}).
Therefore, it is important to establish the presence of a SMB in an
AGN and then investigate the true role of a SMB in  AGN activity.

The most direct 
evidence for a SMB can be obtained by detection of the Kepler orbital
motion of some emission component close to black holes. One technical problem is that 
an expected separation between the two supermassive black holes (e.g.,
$\sim 10^{17}$ cm) is too small to be resolved spatially when we use
optical telescope facilities because 
that separation corresponds to $\sim$100 micro-arcsecond ($\mu$as) 
if its host galaxy is located at a distance of 100 Mparsec (Mpc).
However, the use of  a technique of phase-referencing 
very-long-baseline interferometry (VLBI)
at radio frequencies allows us to achieve such position measurements with 
the accuracy of tens of $\mu$as ({\it 12}). 
We looked for the Kepler motion of a radio-emission
component in a pair of radio sources, 3C 66B and 3C 66A;
3C 66B is a radio galaxy at redshift $z = 0.0215$ ({\it 13}), 
whereas 3C 66A is a more distant
BL Lac object at $z=0.44$ ({\it 14}). This pairing allows us to use this source 
as the stationary position reference to 3C 66B.
They are one of best pairs for the
phase-referencing VLBI, because they are bright radio sources and 
their spatial separation is only 6 arc min.
 
We observed this pair at 2.3 and 8.4 GHz with the very-long-baseline
array (VLBA) of the National Radio Astronomy Observatory (NRAO) over
six epochs between 13 March 2001 and 14 June 2002.
Because the separation between 3C 66A and 3C 66B is smaller than the 
beam widths of the VLBA antennas at 2.3 and 8.4 GHz (26 and 6 arc min,
respectively), it is possible to observe both sources in each antenna
beam simultaneously at these frequencies.

The radio core position of 3C 66B at each epoch was measured with respect to
that of 3C 66A.
We used the local brightest peak in the map, rather 
than the Gaussian peak, to determine the radio core position.
The Gaussian peak may have a larger uncertainty
as a result of the effects of asymmetrical structure of the sources.

Time variations of the radio core position in 3C 66B at 2.3 
and 8.4 GHz (Fig. 1 and Fig. 2, respectively) can be fitted by 
an elliptical motion (Table 1).
The reduced chi-square value obtained when calculated with the observation error is $\approx$ 0.6 at 2.3 GHz and
$\approx$ 0.05 at 8.4 GHz ({\it 12}).
Because these values are less than
unity, especially at 8.4 GHz, 
the position error might have been overestimated. Assuming the reduced chi-square value of unity,  we find that the revised position
errors, $\Delta\alpha'$ and $\Delta\delta'$, are 71 and 60 $\mu$as
at 2.3 GHz and 9 and  7 $\mu$as at 8.4 GHz, respectively ({\it 12}).
Although the orbital periods at 2.3 and 8.4 GHz are nearly the same, 
the major axis at 2.3 GHz is about five times as long as that at 8.4 GHz and 
the position angle of the major axis of the two
frequencies differs by $\approx 24^\circ$. 

The averaged period of the fitted core motion is estimated to be 
$1.05 \pm 0.03$ years.
Although this period might be interpreted as due to some
geodetic effects related to Earth's orbital motion around the sun,
such possibilities can be rejected for the following reasons.
The annual parallax of 3C 66B cannot explain the observed motion, because
the distance of 3C 66B (85 Mpc for the Hubble constant 
$H_0=75$ km s$^{-1}$ Mpc$^{-1}$) leads to the amplitude 
of the annual parallax of only $\sim 0.01$ $\mu$as. 
Furthermore, the large difference in the length of the
major axis between 2.3 and 8.4 GHz can not be understood from our
knowledge of 
the annual parallax. Only a possible variability of the total 
electron content (TEC) in the ionosphere may explain this frequency
dependence, because the excess path due to the ionosphere is
proportional to TEC $\nu^{-2}$, 
where  $\nu$ is the observing frequency ({\it 15}). 
The typical position
error due to the ionosphere for the separation of 6 arc min is
estimated to be  
0.005 $\left(\frac{\nu}{\rm GHz}\right)^{-2}$ $\mu$as,
which is negligible ({\it 16}). 
Another possibility is that the radio core in 3C
66B is subject to microlensing effects by a star in the Milky Way galaxy
near 3C 66B on the celestial plane ({\it 17}) and is
moved by the annual parallax of the star. We attempted to fit the radio
core motion by this microlensing model and obtained the reduced
chi-square value of 1.6, corresponding to a chi-square probability of $\sim
10$ \%. This result indicates that the microlensing model 
does not explain the sinusoidal core motion as well as it explains the elliptical motion
of the radio core.

Because the radio core is located at the root of 
the jet where the optical depth is unity ({\it 18, 19}),
the radio core at 2.3 GHz is expected to be located at a greater distance
{}from the central engine than that at 8.4 GHz. Thus, a precession motion of
the jet provides a natural explanation for the observation that the major axis 
of the radio core orbit is longer at 2.3 GHz than at 8.4 GHz.
Although the orbital major axis should
be always perpendicular to the mean jet axis,
the observed inclination angle between the mean jet axis and the orbital major axis 
is estimated to be 63$\pm 9 ^\circ$ at 2.3 GHz 
and 39$\pm 6 ^\circ$ at 8.4 GHz (calculated on the basis of a $50^\circ$ position angle of the jet,
 which is estimated 
{}from the global jet structure at 2.3 GHz).
This fact indicates that
the observed core motion is not dominated by only the simple precession motion.
The most plausible explanation for this result is that  
the elliptical motion of the radio core is attributable to the combination of the
orbital and precession motions of a SMB in 3C 66B, whereas the
core motion at 2.3 GHz is dominated by the precession motion of
the jet, which at
8.4 GHz is dominated by the orbital motion of the SMB (Fig. 3). 

According to Kepler's third law,
the observed period can be expressed as,

\begin{equation}
 P = 2\pi G^{-1/2}r^{3/2}(M+m)^{-1/2},
\end{equation}
where where $G$ is the gravitation constant, $r$ is the separation of two orbiting black hole, and $M$ and $m$
($M\geq m$) are the masses of the black holes, respectively. Eq. 1
allows us to determine 
the mass density $\rho$ as a function of only $P$,
  \begin{equation}
   \rho\equiv \frac{M+m}{(4/3)\pi r^3} = \frac{3\pi}{GP^2}.
\end{equation}
The observed result leads to $\rho\approx 2\times 10^{15}$
$M_\odot$pc$^{-3}$ (where $M_\odot$ is the mass of the Sun), which is  larger than those of massive dark objects
in NGC 4258 and 
Sgr A$^*$ ({\it 20, 21}). Because the radio core motion
at 8.4 GHz is believed to reflect the orbital motion of 
the SMB more than does that at 2.3 GHz, we 
regard the length of its major axis as the upper limit of the orbital radius of the
SMB. Assuming that the less massive black hole emanates the jet,  we obtain 
the upper limit of the separation and the mass of the SMB of  
$r_{\rm max} \approx 5.4(1+q) \times 10^{16}$ cm and $M_{\rm max}
\approx 4.4 (1+q)^2  
\times 10^{10}M_\odot$, respectively, where $q \equiv\frac{m}{M}$ is the mass ratio
between the objects. We note that these  parameters are 
similar to those of a radio-loud AGN, OJ 287, which is believed to 
be an archetypical AGN with a SMB ({\it 22}). 
This conclusion is consistent with 3C 66B's being a giant elliptical galaxy 
which suggests a galaxy merger.

The SMB in 3C 66B will merge into one as a result of the gravitational radiation
loss. The lifetime in this
evolution phase can be estimated by the relation in Eq. 3 ({\it 23}).

\begin{equation}
 t_{\rm GR} \approx 7.2\times10^4 
  \left(\frac{M}{10^{8}M_{\odot}}\right)^{-3}
  \left(\frac{r}{10^{16}\ {\rm cm}}\right)^4 \frac{1}{q(1+q)}\ \ \ {\rm yr}.
\end{equation}
If we adopt $q=0.1$, we obtain $r_{\rm max} = 5.9 \times 10^{16}$ cm and 
$M_{\rm max} = 5.4 \times 10^{10} M_\odot$. These values give us 
$t_{\rm GR} \simeq 5$ yr for the SMB in 3C 66B. This time scale seems to be 
extremely short compared with a whole life time of a SMB system ({\it
13}). A more reasonable lifetime would be obtained 
for smaller values of $q$ and/or $r$. 
Combining  Eq. 1 and Eq. 3, we find $t_{\rm GR}\propto
r^{-5}$.
If the true $r$ is a quarter of the
corresponding value for the upper limit, a lifetime $\sim 10^3$ times as long
is obtained.  The expected maximum amplitude of gravitational 
radiation from 3C 66B 
will be $\langle h \rangle \sim 10^{-11}$ just before the two black
holes merge into one ({\it 24, 25}). 

\begin{quote}
{\bf References and Notes}

\begin{enumerate}
\item  A. Sillanp${\rm \ddot{a}\ddot{a}}$, S. Haarala, 
 M. J. Valtonen, B. Sundelius,   G. G.  Byrd,  {\it Astrophys. J.} {\bf 325,} 628 (1988).
\item { E. Valtaoja,  {\it et al.,} {\it Astrophys. J.}
 {\bf 531,} 744 (2000).} 
\item { N. Roos,  J. S. Kaastra,   C. A.  Hummel,  {\it
 Astrophys. J.} {\bf 409,} 130 (1993).}
\item { Z. Abraham,  {\it Astron. Astrophys.} {\bf 355,} 915 (2000).}
\item { D.  Merritt, R. D.  Ekers, {\it Science} {\bf 297,}
 1310 (2002).}
\item { C. M. Gaskell, in Jets from Stars and Galactic Nuclei,
W. Kundt, Ed.  (Springer-Verlag, New York, 2002), p 165}
\item { D. B. Sanders,  {\it et al.,} {\it Astrophys. J.} {\bf 325,}
 74 (1988).}
\item { Y. Taniguchi, S. Ikeuchi,  Y.  Shioya, {\it
 Astrophys. J.} {\bf 514,} L9 (1999).} 
\item { J.  Kormendy,  D. Richstone,  {\it Annu. Rev.} {\it
 Astron. Astrophys.} {\bf 33,} 581 (1995).} 
\item { M. J.  Rees, {\it Annu. Rev.} {\it Astron. Astrophys.}
 {\bf 22,} 471 (1984).} 
\item { M. C. Begelman,  R. D. Blandford, M. J.   Rees,
 {\it Nature} {\bf 287,} 307 (1980).} 
\item { Materials and Methods are available as supporting material
in {\it Science} Online. }
\item { T. A. Matthews, W. W. Morgan, M.   Schmidt, {\it Astrophys. J.}
 {\bf 140,} 35 (1964).}
\item {  H. R. Miller,  B. Q. McGimsey,  {\it
 Astrophys. J.} {\bf 220,} 19 (1978).} 
\item {  A. R. Thompson, J. M. Moran, G. W. Swenson, 
in {Interferometry and Synthesis in Radio Astronomy}, (Wiley, New York, 1986) p 554} 
\item {  The excess path in the zenith direction due to
 ionosphere is $
  l(\nu) = {40.3}~{\rm TEC}\ {\nu^{-2}}
 $ ({\it 15}). As a result of  the Doppler shift by the
 Earth's spin, a
frequency shift between two sources occurs, given by  $\nu' = \nu(1 +  
\omega \theta D/c)$, where $D$ is the baseline length, and
$\omega$ is the 
angular velocity of the 
Earth spin. Thus, dividing $ l(\nu) - l(\nu')$ by $c$, we obtain the time
delay due to ionospheric excess path. We assumed that TEC in the zenith direction
 is $6\times 10^{17}$ m$^{-2}$. } 
\item {  S. Refsdal,  {\it
Mon. Not. R. Astron. Soc.} {\bf 128,} 295 (1964).} 
\item { A. K$\ddot{\rm o}$nigl, {\it Astrophys. J.} {\bf 243,} 700 (1981).}
\item { A. P. Lobanov,  {\it Astron. Astrophys.} {\bf 330,} 79 (1998).}
\item { M. Miyoshi,  {\it et al.,}  {\it Nature}
 {\bf 373,} 127 (1995).} 
\item { A. M. Ghez, M. Morris, E. E. Becklin, A. Tanner,   
  T.  Kremenek, {\it Nature} {\bf 407,} 349 (2000).}
 \item { H. J. Lehto,  M. J.  Valtonen,   {\it Astrophys. J.} {\bf 460,}
 207 (1996)} 
\item {23. Q. Yu, {\it
 Mon. Not. R. Astron. Soc.} {\bf 331,} 935 (2002).} 
\item { K. P.  Thorne,  V. B.  Braginsky,  {\it
 Astrophys. J.} {\bf 204,} L1 (1976).} 
\item { T. Fukushige,   T. Ebisuzaki,  {\it
 Astrophys. J.} {\bf 396,} L61 (1992)} 
 \item { Supported by a
grant-in-aid for Fellows of the Japan Society for the Promotion of Science from the Japanese Ministry of
Education, Culture, Sports, and Science. NRAO is a facility of the
 National Science Foundation, operated under 
cooperative agreement by Associated Universities, Inc.
 We acknowledge anonymous reviewers for useful
 suggestions. }
\end{enumerate}
\end{quote}

\newpage

\begin{table}
\caption{Parameters of the fitted orbital motion of the radio core in 3C 
 66B. 
 The errors in the parameters correspond to a change of 1 in
 chi-square from the value at the best fitted parameter. The offset angle is the angle between the major axis and the point where the
 orbital motion started.  
 R.A. center is the relative position in right ascension.
 Dec. center is the relative position in declination.}
 \begin{tabular}{lccc}
  \hline
  
{} &
{} &
2.3 GHz&
8.4 GHz\\
  \hline
  Major axis &($\mu$as)& 243$\pm$30 & 45$\pm$4 \\
  Axial ratio &   & 0.31$\pm$0.17 & 0.24$\pm$0.14 \\
  Orbital period &(yr)& 1.10$\pm$0.06 & 1.02$\pm$0.04 \\
  Position angle &($^\circ$)& 113$\pm$9 & 89$\pm$6 \\
  Offset Angle  &($^\circ$) & 60$\pm$7 & 101$\pm$5 \\
  R.A. center &(mas)   & 1.441$\pm$0.020 & 0.970$\pm$0.002 \\
  Dec. center  &(mas)  & $-$0.888$\pm$0.033 & $-$1.861$\pm$0.004 \\
\hline
 \end{tabular}
 
 \end{table}

\bibliography{scibib}

\bibliographystyle{Science}

\noindent {\bf Fig. 1.} Orbital fit calculations applied to the data of the position 
change of the core at 2.3 GHz. Observations were carried out on 13
March, 25 June, and 9 November 2001, and 8 February, 21 February, and 14 June
2002.  We show ({\bf A}) spatial distribution, ({\bf B}) time 
evolution toward right ascension direction, and ({\bf C}) time
evolution toward declination direction. The error bars indicate the
revised position errors suggested from the chi-square value; 71
and 60 $\mu$as 
at 2.3 GHz and 9 and  7 $\mu$as at 8.4 GHz, in right ascension
and declination, respectively ({\it 12}).

\noindent {\bf Fig. 2} Same as Fig. 1, but at 8.4 GHz. 

\noindent {\bf Fig. 3} A schematic view of the geometrical relation between
the radio core motion and the SMB. Because the orbital velocity 
of the SMB is added to the intrinsic jet velocity,
the jet precession can be induced with the same period as that of the SMB
orbital motion ({\it 3}).  Therefore, 
if the radio core is far enough from the SMB,
the orbital major axis of the radio core should be observed as
perpendicular to the mean jet axis, as is the normal precession. 
However, if the radio core is very close to the SMB,
it should be observed as almost parallel to the orbital plane of the SMB. 
Therefore, the position angle of the radio core orbit depends on 
the distance between the radio core and the SMB, 
if the jet direction misaligns with the angular momentum  
vector of the SMB orbit.

\end{document}